\documentclass[12pt]{spieman}
\usepackage[utf8]{inputenc}
\usepackage{amsmath,amsfonts,amssymb}
\usepackage{graphicx}
\usepackage{setspace}
\usepackage{tocloft}

\usepackage{multirow}
\usepackage{xcolor}
\usepackage{soul}
\usepackage{newtxmath}
\usepackage{orcidlink}

\newcommand{\um}{$\upmu$m}

\title{Fat Cosmic Ray Tracks in CCDs}
\author[a,b]{Theodore A. Grosson\orcidlink{0000-0002-5729-8894}}
\author[b]{Andrei Nomerotski\orcidlink{0000-0001-5444-5345}}
\author[ ]{The LSST Dark Energy Science Collaboration}
\affil[a]{University of Victoria, Department of Physics and Astronomy, Victoria, British Columbia, Canada}
\affil[b]{Brookhaven National Laboratory, Physics Department, Upton, NY, United States}

\graphicspath{{img/}}

\begin{document}

\maketitle

\begin{abstract}
    Cosmic rays are particles from the upper atmosphere which often leave bright spots and trails in images from telescope CCDs. We investigate so-called ``fat" cosmic rays seen in images from Vera C. Rubin Observatory and the Subaru Telescope. These tracks are much wider and brighter than typical cosmic ray tracks, and therefore are more capable of obscuring data in science images. By understanding the origins of these tracks, we can better ensure that they do not interfere with on-sky data. We compare the properties of these tracks to simulated and theoretical models in order to identify both the particles causing these tracks as well as the reason for their excess spread. We propose that the origin of these tracks is cosmic ray protons, which deposit much greater charge in the CCDs than typical cosmic rays due to their lower velocities. The generated charges then repel each other while drifting through the detector, resulting in a track which is much wider than typical tracks.
\end{abstract}


\section{Introduction}
\label{sec:intro}

Cosmic rays create difficulties for astronomers since it is impossible to protect telescopes from them. Highly energetic particles incident on Earth's atmosphere create secondary particles which leave tracks in CCD images, the most common particles being muons \cite{ParticleReview}. These tracks appear practically in each image with long enough exposure, and they must be removed before useful analysis can occur.

While CCDs detect incident photons through the photoelectric effect, with generated electrons drifting through the silicon until they are collected at the gates, charged particles such as cosmic rays deposit energy in the detectors through ionization and excitation of electron-hole pairs in the silicon. The energy deposition depends on the characteristics of the incident particle, so the images they create can reveal information about the particle which generated them \cite{particle-detectors}.

Typically, due to the high momentum of incident muons, the tracks appear straight and narrow \cite{Merlin}. If multiple exposures of a target are obtained, the tracks can be removed simply by observing the difference between exposures. If only one exposure is available, though, it is necessary to identify cosmic rays through their morphology; in particular, the muon tracks can be identified by their sharper edges. These tracks are smeared only due to the processes internal to silicon, such as diffusion, without contribution from the telescope  optics nor the atmosphere, and, therefore, are sharper than typical star images \cite{Groom-CRs}. The sharpness of star images is defined by the point spread function (PSF) with contributions from the telescope optics and atmosphere.

\begin{figure}[t]
    \centering
    \includegraphics[width=0.8\textwidth]{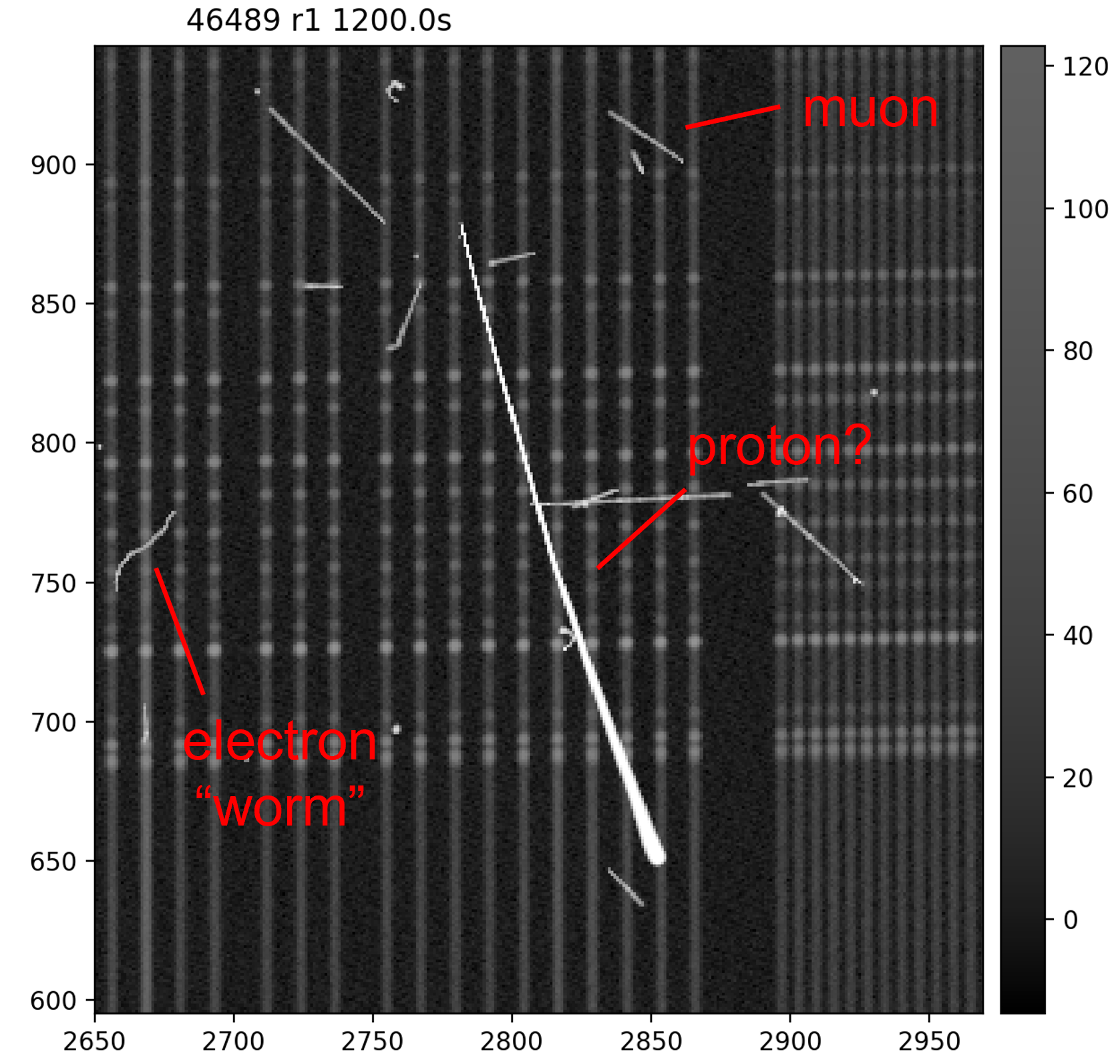}
    \caption{1200 second dark exposure from PFS at the Subaru Telescope featuring cosmic rays, including the bright ``fat" track in the center. Image from Robert Lupton and the PFS Collaboration\cite{PFSimg}.}
    \label{fig:fat}
\end{figure}

However, one type of track which was noticed in dark exposures from the Prime Focus Spectrograph (PFS), referred to here as a ``fat" cosmic ray, is much brighter and wider than typical muon tracks, as seen in Figure \ref{fig:fat}. These fat tracks may be difficult to identify and remove using morphology-based cosmic ray detection. In this study, we investigate the properties of fat tracks in order to identify their cause. We compare these measurements to simulations of cosmic rays in CCDs, as well as to expected abundances of different particles. Understanding the origins of these fat tracks could enable astronomers to devise the best ways to ensure they do not interfere with on-sky data.

Our analysis suggests that protons are the most likely source of these tracks. For slow particles, energy deposition into CCDs follows $\beta^{-2}$ (the Bethe-Bloch equation \cite{ParticleReview}), so fat tracks must necessarily be caused by particles significantly slower than typical cosmic rays. At sea level, this is true of protons, which have nine times the mass of muons but lower mean kinetic energy, i.e., significantly lower mean velocity. In addition, strongly interacting cosmic ray protons become less common at lower altitudes as they interact with the atmosphere and create secondary cosmic rays, such as muons \cite{ParticleReview}. This agrees with our observations of lower rates of fat tracks at lower altitudes.

Large energy deposition in silicon sensors has been studied before. One example is a phenomenon known as the ``plasma effect," which has been researched in the context of interactions of  $\alpha$\nobreakdash-particles in semiconductor detectors. The $\alpha$\nobreakdash-particles have been seen to leave large, round spots of ionization due to the deposition of charge dense enough to satisfy the plasma condition \cite{plasma-effect1,plasma-effect2,plasma-effect3,SNOLAB-alphas}. One consequence of this effect is that the high charge density repels charge carriers away from the initial path of the particle, which is the mechanism by which we believe fat tracks are produced. However, typical $\alpha$-particle energies (a few MeV) result in the production of ${\sim} 10^6$ electron/hole pairs in a compact volume of few pixels. We note that in our case of fat tracks the ionization density is equal to about $10^4$ electrons per pixel, which is at least an order of magnitude lower than the ionization from alphas described above.

Other studies report the appearance of protons in a silicon pixel detector for radiation imaging \cite{proton-track-angles, timepix}. The authors find tracks that look very similar to our own tracks when the protons are incident to the detector at large angles, supporting our proton origin theory for fat tracks.

In Section \ref{sec:spread}, we describe how we measure observed cosmic rays and compare them to simulations of cosmic rays in CCDs, and we discuss the method by which high energy deposition results in wider than usual cosmic ray tracks. Section \ref{sec:particles} examines the particles which cause fat tracks, and Section \ref{sec:conclusion} states our conclusions on the origin of fat tracks.

\section{Identifying cause of excess spread}
\label{sec:spread}

\subsection{CCDs and data used}

Images used in the analysis in Section \ref{sec:spread} are 300-second dark exposures taken with the Vera C. Rubin Observatory Commissioning Camera (ComCam) at the facility on Cerro Pach\'{o}n, Chile. ComCam consists of nine 4k x 4k CCDs which are each 100 \um{} thick and have a pitch of 10 \um{} \cite{ComCam}. The images have been bias-subtracted and converted to units of electrons from ADU. Cosmic rays were identified in these images using the Legacy Survey of Space and Time\footnote{\url{https://rubinobservatory.org/}} (LSST) Science Pipelines\footnote{\url{https://pipelines.lsst.io.}} \cite{pipelines}, and cutouts of the images containing one track each were extracted and saved in individual FITS files. An example is shown in Figure \ref{fig:process}(a). The cutouts were selected so that the cosmic ray track traverses the entire diagonal of the cutout.

Additional dark exposures used in Section \ref{sec:particles} were taken with the Rubin Observatory LSST Camera at SLAC National Accelerator Laboratory and the red-sensitive arm of the Prime Focus Spectrograph at the Subaru Telescope. The LSST Camera consists of 189 CCDs of two types with about a half of them identical to those in ComCam \cite{LSST}, and PFS uses two 2k x 4k, 200 \um{} thick, 15 \um{} pitch CCDs in its red-sensitive arm \cite{PFS}. In all cases the CCD sensors were fully depleted n-channel detectors.

\subsection{Measuring track widths}

Following the example of Ref.~\citenum{Merlin}, we select only linear tracks by rejecting any cutouts with a linear correlation coefficient ($R^2$) less than 0.95. This has the effect of filtering out ``worms," which are low-energy electrons that leave highly curved tracks. The precise length of the tracks can then be obtained by fitting a line to the tracks. We first fit an estimate line to pixels whose value is greater than 30 electrons. To remove any artifacts in the cutouts which are not part of the main cosmic ray track, such as $\delta$ ray electrons, we mask all pixels which are farther than 1.5 standard deviations from the best-fit line, or brighter than 5 standard deviations from the mean. We then fit a new line which ignores the masked pixels and is weighted by the pixel values.

The distance the cosmic ray travelled through the pixels at each end of the track can be determined by the pixel values, which scale with the distance travelled through those pixels. We calculate the average energy deposition rate $dE/dx$ of the track in e\textsuperscript{\textendash} pixel$^{-1}$ by summing the pixel values of the track and dividing by the length of the best fit line, excluding the terminal pixels. The brightness of the terminal pixels in e$^-$ divided by $dE/dx$ is the distance the cosmic ray travelled through those pixels, giving us the endpoints of the best-fit line and thus a more precise track length and $dE/dx$ value.

\begin{figure}[t]
    \centering
    \includegraphics[width=0.9\textwidth]{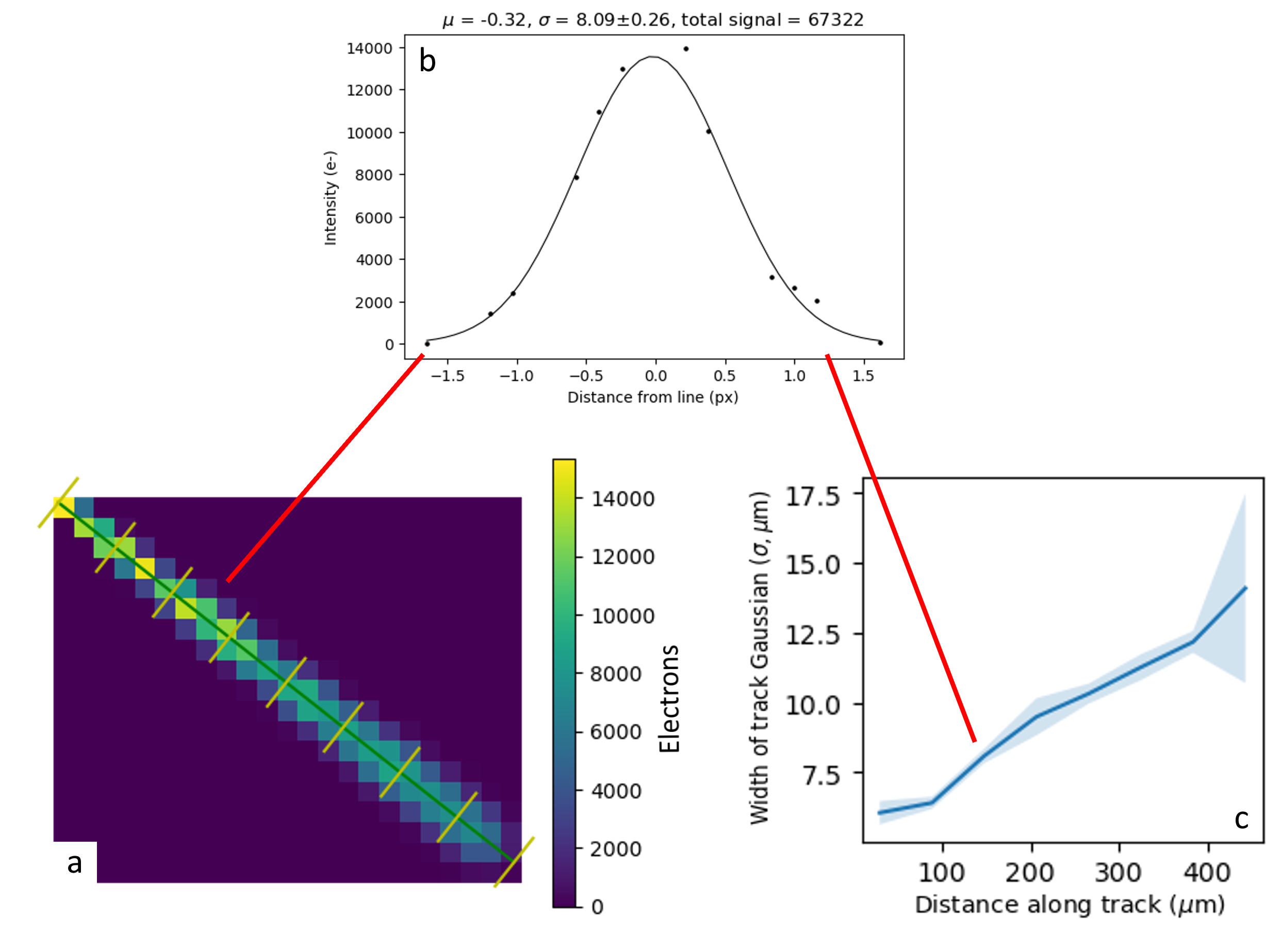}
    \caption{Process of measuring the width of a fat track. Panel a: the track's best fit line and segment divisions. Panel b: the best fit Gaussian for the pixels of one segment. Panel c: the measured width of the track along its length.}
    \label{fig:process}
\end{figure}

In order to determine how the width of fat tracks behaves compared to regular muon tracks, we divide each track into eight equally-sized segments and measure the width of the track in each segment. This is done by fitting a Gaussian to the pixel values as a function of their distance to the center of the track; the width of the track in each segment is taken to be the value of $\sigma$ for that segment's best-fitting Gaussian. This process is shown in Figure \ref{fig:process}.

With the width of each segment in hand, we can characterize how the width of each track changes over its length. Since we assume that all our tracks traverse the full thickness of the CCD, we can assign an average height to each track segment. Again following Ref.~\citenum{Merlin}, we can take a selection of tracks and find the average width at each height, which gives us the point-spread function (PSF) for that collection of tracks.

After charges are generated in a CCD by, for example, a cosmic ray, the charges drift through the thickness of the CCD until they reach the gate, after which they can be measured. While the charges are drifting, they tend to diffuse away from the location at which they were generated, resulting in wider spots for charges which are generated far from the gate. Thus the width of a cosmic ray track should increase as it passes through the thickness of the CCD \cite{Merlin,SNOLAB-PSF}. The PSF of muons is approximately linear with height, with a component from this diffusion plus a constant from the intrinsic resolution of the CCD, which depends on the size of the pixels. The expected PSF $\sigma$ can be calculated from the equation\cite{Merlin} \[\sigma^2(\Delta z) = 2 \frac{kT}{e} \frac{D \Delta z}{V} + \frac{pixel~pitch^2}{12}\] where $\Delta z$ is the height at which the charge is generated, $D$ is the sensor thickness, and $V$ is the bias voltage. In the case of ComCam, $V=40$ V and $T \approx -100$C.

\begin{figure}[t]
    \centering
    \includegraphics[width=0.8\textwidth]{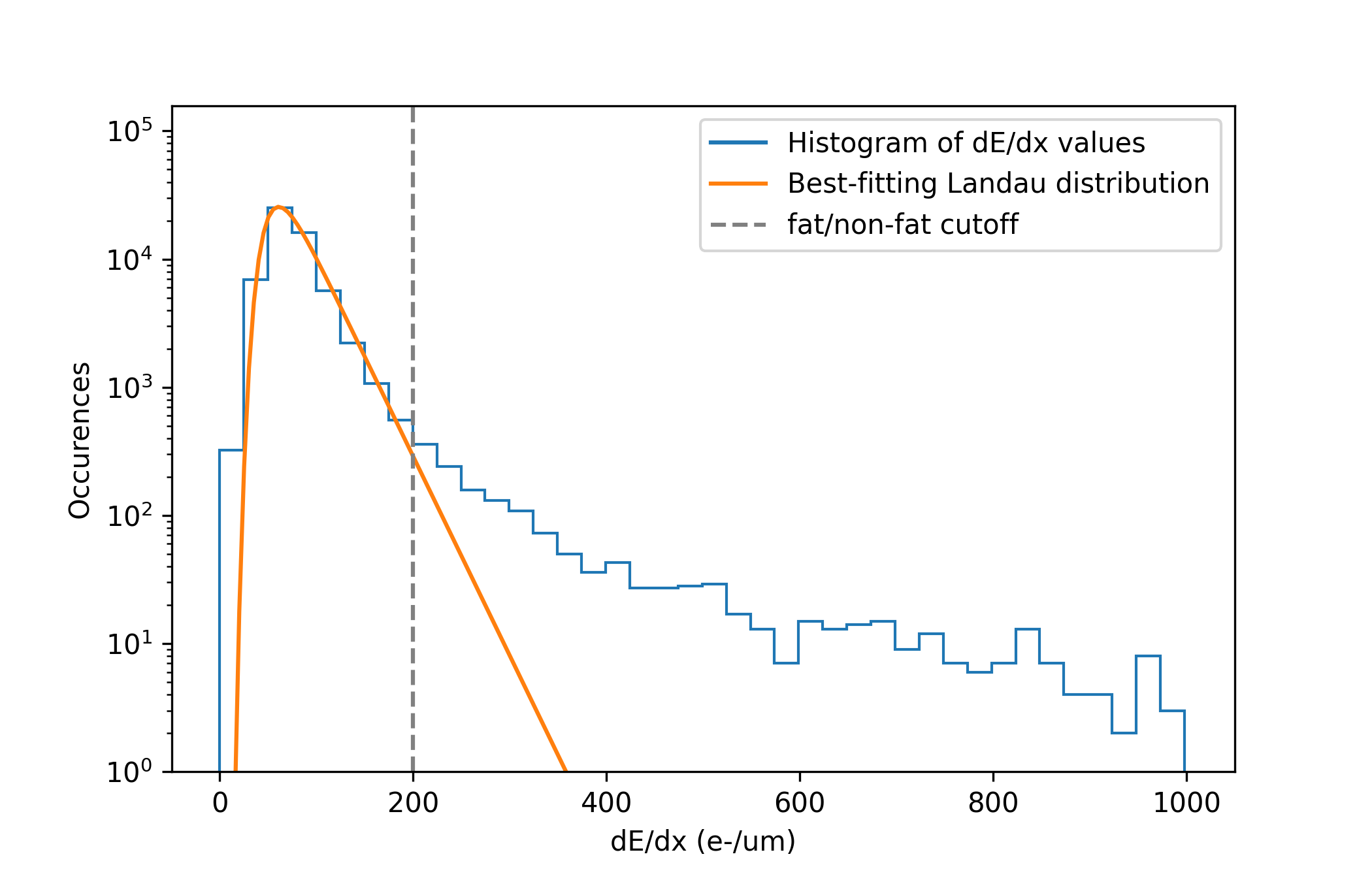}
    \caption{Distribution of $dE/dx$ for track segments below 1000 e$^-$ \um$^{-1}$, and the best fitting Landau function for this distribution. Segments brighter than 1000 e$^-$ \um$^{-1}$ account for less than 0.2\% of the sample.}
    \label{fig:dedxs}
\end{figure}

\begin{figure}[ht]
    \centering
    \includegraphics[width=0.8\textwidth]{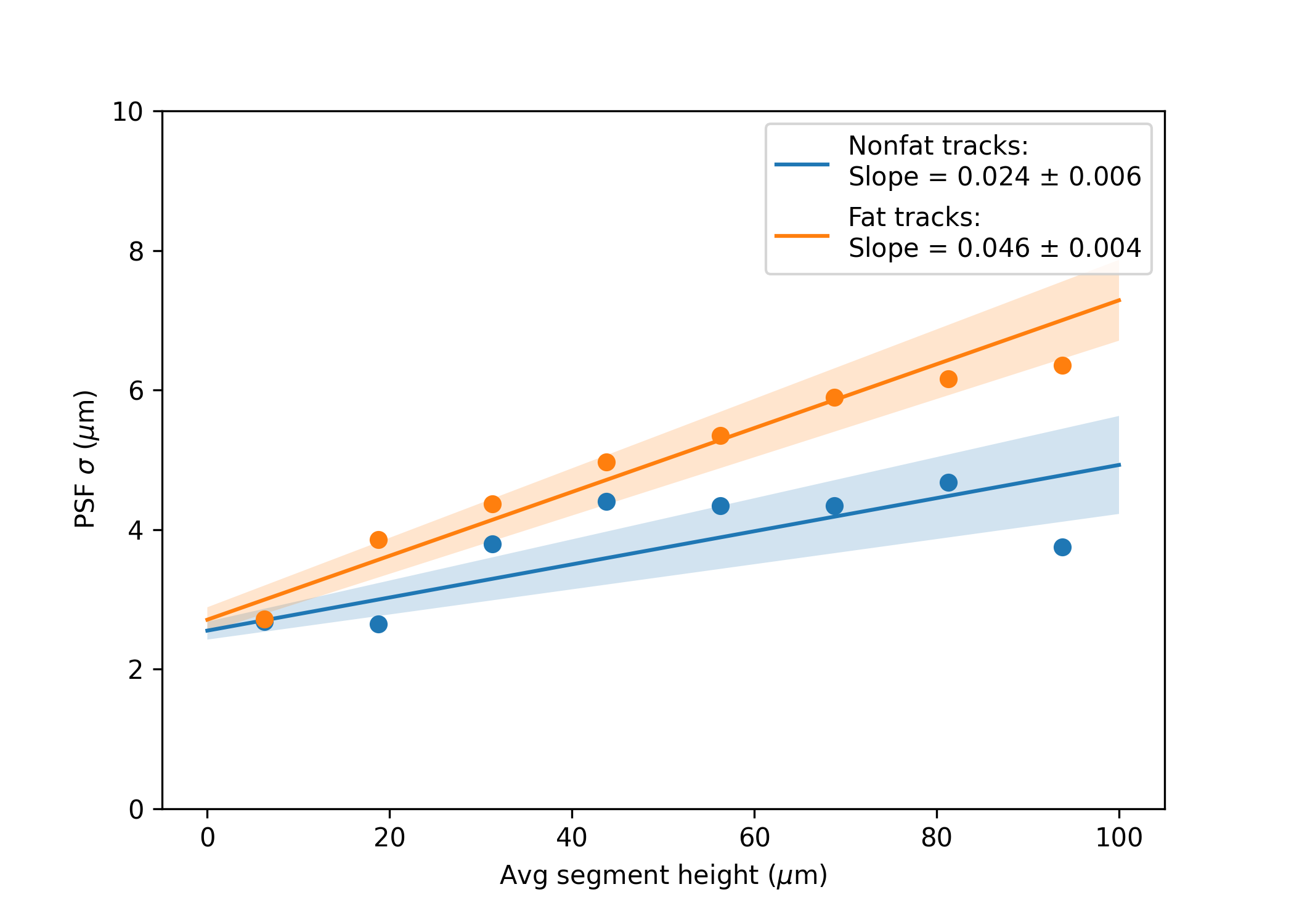}
    \caption{PSFs for fat and non-fat tracks. Shaded regions indicate uncertainty of the PSF best fit. The expected resolution for this detector, corresponding to the intercept of the PSF, is 2.9 \um.}
    \label{fig:PSFs}
\end{figure}

We divide our cutouts into fat and non-fat groups by defining a cutoff $dE/dx$ value: tracks with an average $dE/dx > 200$ e$^-$ \um$^{-1}$ are defined as fat. The distribution of measured $dE/dx$ values less than 1000 e$^-$ \um$^{-1}$ is shown in Figure \ref{fig:dedxs}. The distribution of energy deposition rates is described by a Landau distribution, so the best-fitting Landau distribution is also plotted \cite{ParticleReview}. Our cutoff $dE/dx$ is right where the histogram separates from the distribution function. We then compare the PSFs of the two groups, which are shown in Figure \ref{fig:PSFs}. We find that the intercept, which corresponds to the intrinsic resolution of the CCD, remains constant for both types of tracks, as we expect for tracks coming from the same detector. The intercept is comparable to the expected resolution for ComCam, which is 2.9 \um{}. However, the greater slope of the fat tracks' PSF indicates a significant spread in excess of typical diffusion in fat tracks.

Since we have identified additional broadening of fat tracks on top of diffusion, there is an ambiguity regarding the tracks: the second broadening effect could add to diffusion, increasing the slope of the PSF; or it could oppose it, decreasing the slope. In the latter case, it is possible the second effect could be strong enough to make the slope of the PSF negative, such that the tracks become wider closer to the gate. However, because we identify the height of each segment based on its width, we are unable to detect whether this is the case.

Despite this ambiguity, we are unable to find any significant variation of the PSF slopes of individual fat tracks that would correspond to these two options. In addition, we find that our data is much more well-fit by the simulations described in Section \ref{sec:simuations} if we assume that the wider ends of the tracks are indeed farther from the gates.

\subsection{Simulating spread}
\label{sec:simuations}

We propose that the excess spread is caused by an effect similar to the brighter-fatter effect described by Ref.~\citenum{BF}, in that electrostatic repulsion within the sensor is responsible for additional width, and this width scales with the brightness of the source. The brighter-fatter effect describes how the accumulation of charges in CCD pixels distorts the electric field within the sensor, pushing incoming charges away from overfilled pixels. Unlike the brighter-fatter effect, in this case large numbers of near-instantaneously generated electrons create significant repulsion between themselves as they drift towards the gates, resulting in a more dispersed incoming cloud of electrons. This effect becomes noticeable for charge deposition rates greater than $\sim$few hundred electrons per micron.

\begin{figure}[t]
    \centering
    \includegraphics[width=0.95\textwidth]{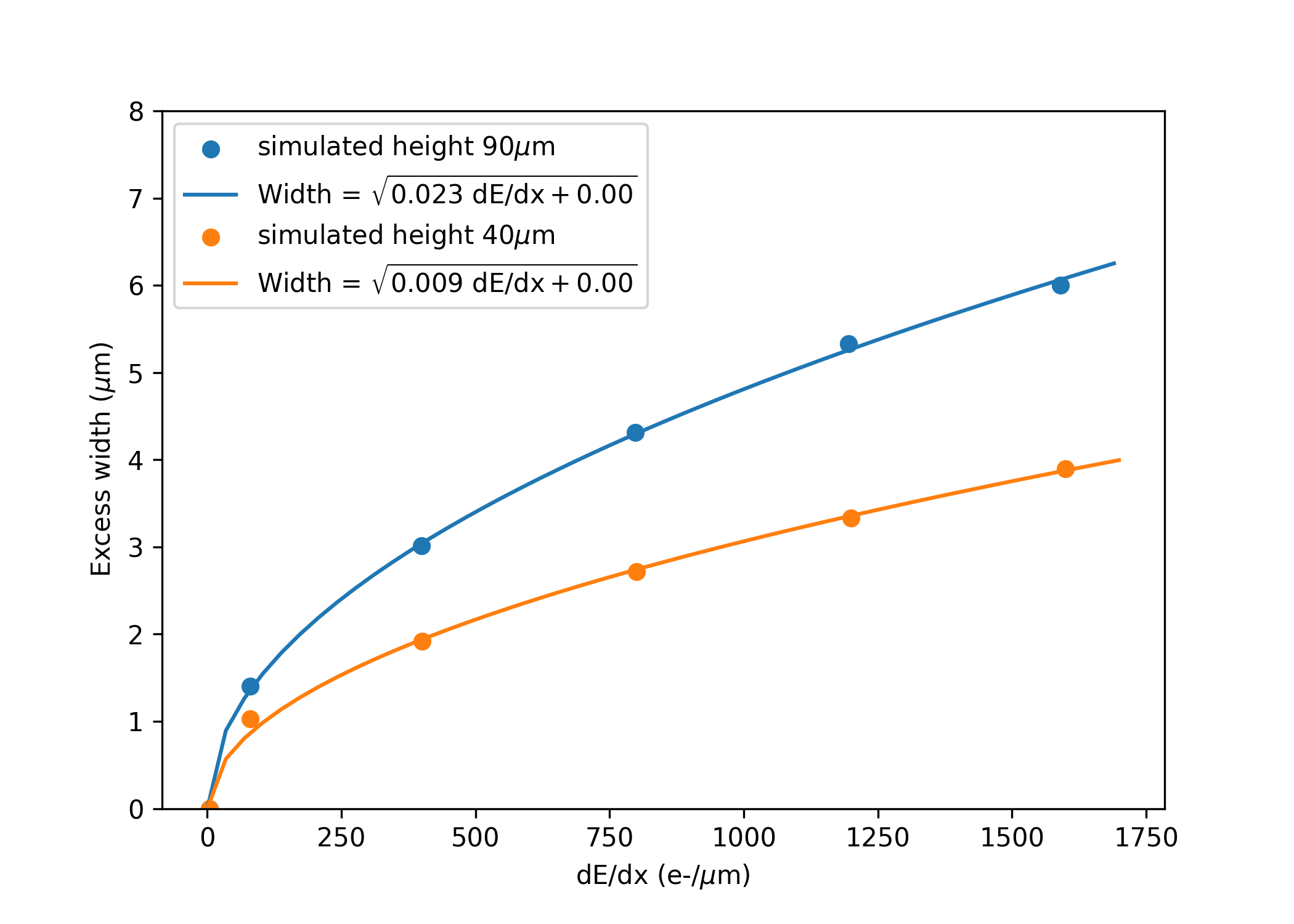}
    \caption{Excess widths of simulated cosmic ray tracks for varying $dE/dx$ and heights, and the best-fitting square root function for each height. Widths here are after subtraction of simulated diffusion and resolution components.}
    \label{fig:sim_width}
\end{figure}

To test this explanation, we model cosmic ray events within a CCD using the CCD simulator developed by Ref.~\citenum{Poisson_CCD}\footnote{https://github.com/craiglagegit/Poisson\_CCD}. This code numerically solves Poisson's equation and calculates charge transport through the CCD, accounting for both diffusion and electrostatic effects between charges. We model a 7x7 grid of pixels in a detector with the same configuration as ComCam \cite{ComCam}, then model a cosmic ray event as a line of electrons with a constant linear density ($dE/dx$), constant height above the gates, and aligned with the pixel grid so that the line is directly above the center pixel. We perform 12 simulations, varying $dE/dx$ from 4 to 1600 e\textsuperscript{-} \um$^{-1}$ at a height of either 40 or 90 \um{} above the gates. The resulting image is stored as an array of electron counts for each pixel. The width of the simulated track is calculated the same way as a single segment of the observed cosmic ray tracks, and the diffusion and resolution components are subtracted in quadrature to leave only the excess widths. The diffusion and resolution components are taken to be the simulated widths at $dE/dx=4$ e\textsuperscript{-} \um{}$^{-1}$. These measured excess widths are shown in Figure \ref{fig:sim_width}.

Ref.~\citenum{repulsion} shows that electrostatic repulsion is significant for spherically symmetric charge packets in silicon drift detectors somewhat thicker than our CCDs (300 compared to 100 \um{}), finding that the radius of a cloud of charges $r\propto(Nt)^{1/3}$, where $N$ is the number of charges and $t$ is the elapsed time. Considering the two-dimensional case, as would apply to our initial line of charge, has the effect of creating a square root dependence on linear charge density. We fit our simulations at each height to an equation of the form $Excess~width = \sqrt{a~dE/dx+b}$ to demonstrate the effect of varying $dE/dx$.

The excess width of the track also has a height dependence in addition to that expected from diffusion. This is expected, as charges which begin farther from the gate experience repulsion over a greater amount of time.

\subsection{Simulations vs observations}

\begin{figure}[ht]
    \centering
    \includegraphics[width=0.9\textwidth]{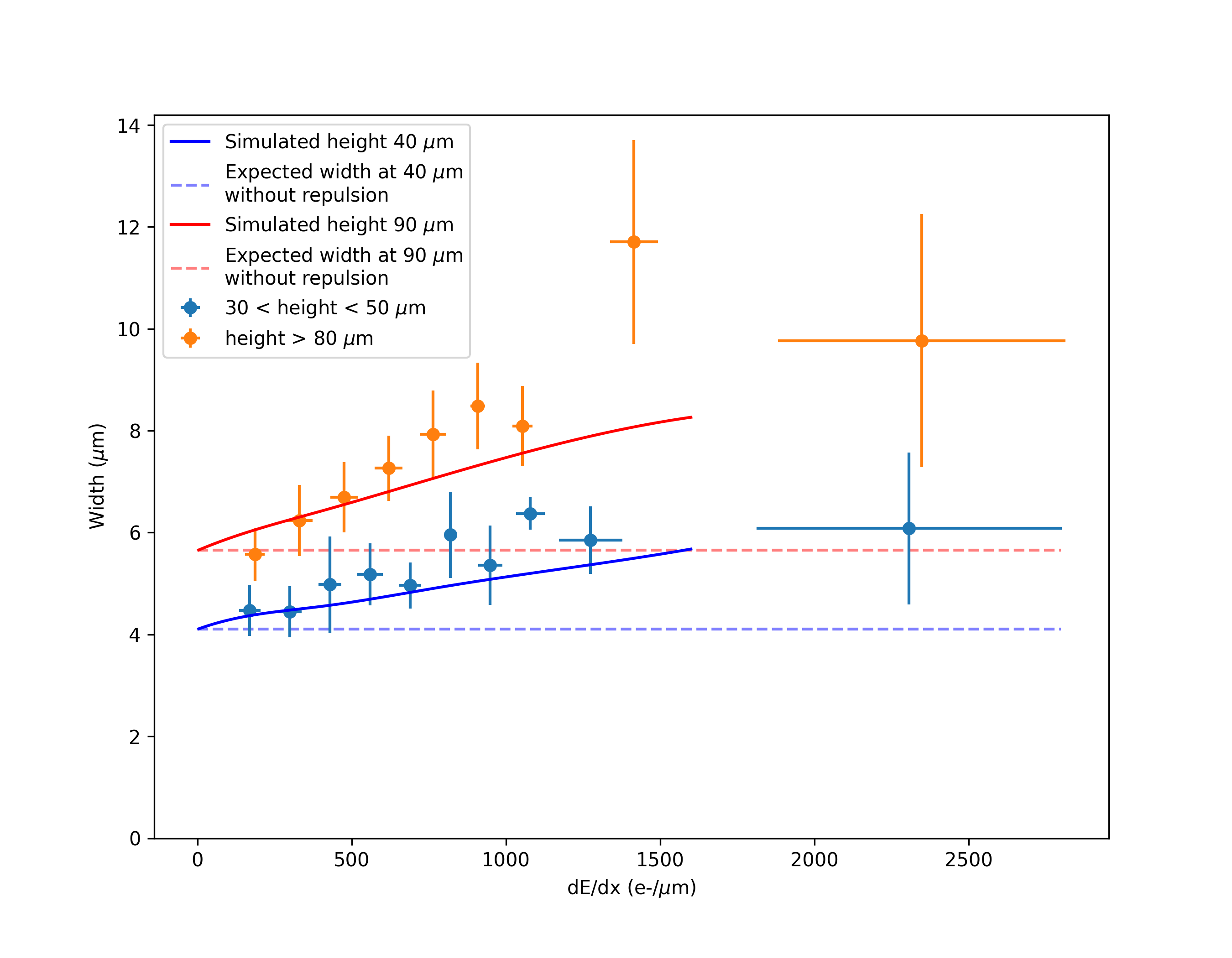}
    \caption{Width vs brightness for fat tracks (binned by $dE/dx$) and cubic spline fit to simulated tracks at different heights. Diffusion and resolution components are included.}
    \label{fig:width-dedx}
\end{figure}

Figure \ref{fig:width-dedx} shows the simulated track widths, with diffusion and resolution components still included, alongside the measured widths of fat track segments. To illustrate the height dependence on width, we selected two groups of track segments: those with height between 30 and 50 \um{} and those with height greater than 80 \um{} from the CCD gates. These groups were then binned by $dE/dx$, with errorbars showing the standard deviations of both the width and $dE/dx$ values within the bins.

Figure \ref{fig:width-dedx} shows good agreement between observations and simulations. Using a cubic spline to interpolate between simulated points, we calculate reduced chi-squared $\chi_{\rm{red}}^2 = 1.44$, suggesting that repulsion between electrons within the sensor is the primary mechanism for track broadening.

\section{Identifying particles}
\label{sec:particles}

In addition to identifying the mechanism by which cosmic ray tracks are broadened, we also wish to identify the particles which cause fat tracks. Since $dE/dx$ of cosmic rays in CCDs scales as $\beta^{-2}$, fat tracks must have relatively low velocities to obtain their brightness. One particle which fits this requirement is cosmic ray protons. Muon cosmic rays (mass 106 MeV) have a mean energy of 4 GeV at sea level, with $\beta << 1$ muons rare \cite{ParticleReview}. Meanwhile, cosmic ray protons (mass 938 MeV) have a mean energy of 1 GeV at sea level, and so typically have much lower $\beta$ than muons \cite{CRs}.

\begin{figure}[t]
    \centering
    \includegraphics[width=0.8\textwidth]{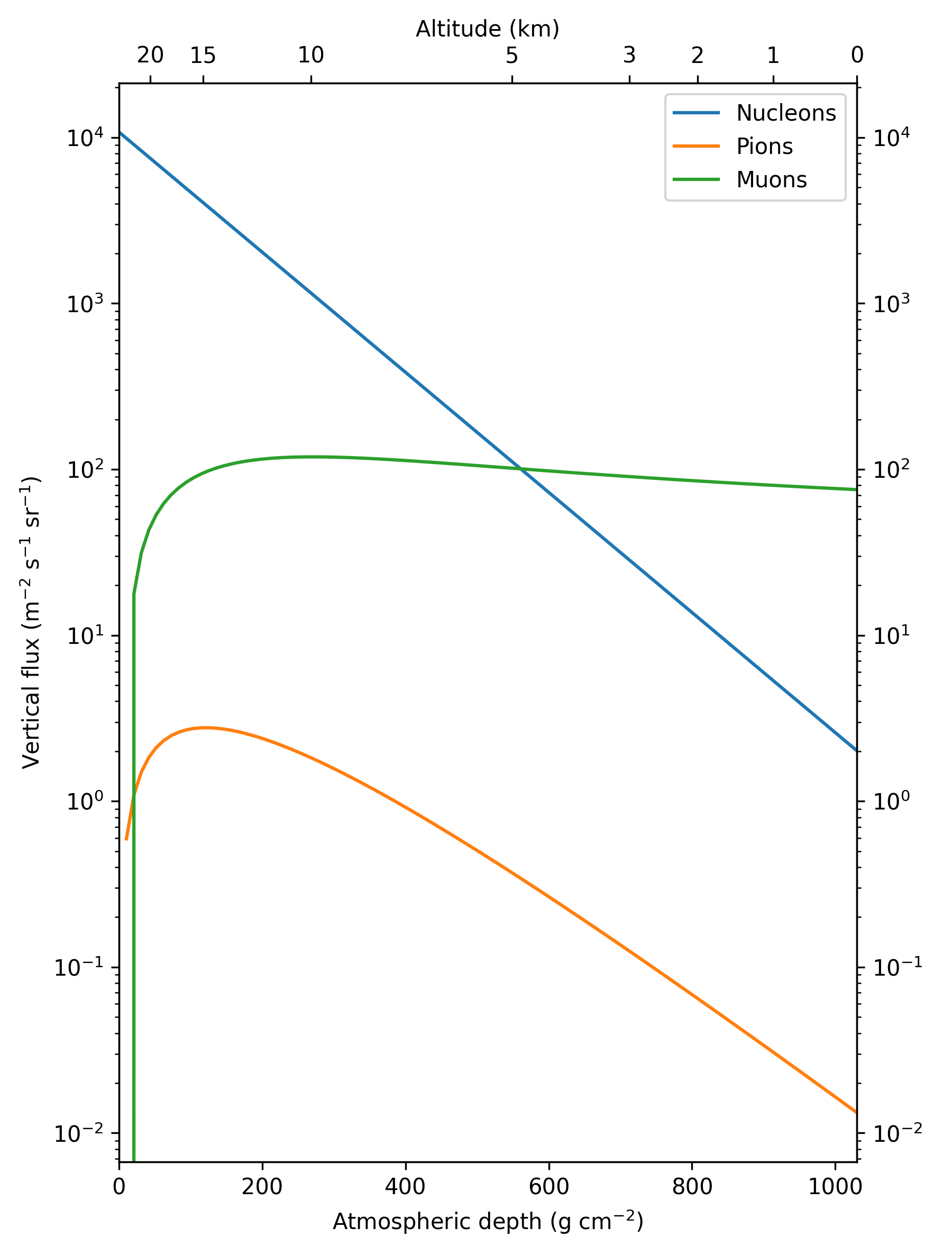}
    \caption{Estimates of vertical fluxes of cosmic rays in the atmosphere with $E>1$ GeV. Based on Figure 30.5 from Ref.~\citenum{ParticleReview}.}
    \label{fig:altitude}
\end{figure}

Fat tracks were not observed in images from Rubin Observatory CCDs while being developed at Brook\-haven National Laboratory (although a systematic search has not been performed); they were only first noticed once images had been taken at higher altitudes. Protons match this observation as well, as their interactions with the atmosphere result in the creation of secondary particles such as pions, which then decay to muons. At lower altitudes, protons become less common as they produce more secondary particles. On the other hand, muons do not interact strongly with the atmosphere, and therefore many of them reach the surface \cite{ParticleReview}. The fluxes of muons and protons in the atmosphere as a function of altitude can be approximated using methods described by Refs.~\citenum{CR-fluxes,muon-fluxes}. The calculated fluxes of cosmic ray nucleons, muons and pions at different altitudes using these methods is shown in Figure \ref{fig:altitude}.

To compare expectations of proton and muon abundances to observations, we analyze dark images from three different cameras: Prime Focus Spectrograph at the Mauna Kea Observatory; ComCam at Rubin Observatory; and the LSST Camera, located at SLAC National Accelerator Laboratory during data collection. Cosmic rays were extracted from images and fat tracks were identified using the same methods described above. The number of tracks identified in each camera, along with the cameras' altitudes, are listed in Table \ref{tab:altitudes}. Note that in the case of PFS at Mauna Kea, the statistics are limited and, therefore, the uncertainty is large.

\begin{table}[ht]
    \centering
    \begin{tabular}{|l|c|c|c|}
        \hline
        \textbf{Camera /} & \multicolumn{1}{|l|}{\textbf{PFS /}} & \multicolumn{1}{|l|}{\textbf{ComCam /}} & \multicolumn{1}{|l|}{\textbf{LSST Camera /}} \\
        \textbf{Location} & \multicolumn{1}{|l|}{\textbf{Mauna Kea}} & \multicolumn{1}{|l|}{\textbf{Cerro Pach\'{o}n}} & \multicolumn{1}{|l|}{\textbf{SLAC}} \\
        \hline
        Altitude & 4.14 km & 2.66 km & 0.1 km \\
        \hline
        Total tracks & 406 & 29,149 & 34,350 \\
        \hline
        Ratio fat tracks/ & \multirow{2}{*}{0.5$\pm$0.3\%} & \multirow{2}{*}{0.18$\pm$0.02\%} & \multirow{2}{*}{0.050$\pm$0.01\%} \\
        total tracks &  &  &  \\
        \hline
        Expected ratio & $\sim$10x sea level & $\sim$5x sea level & (sea level) \\
        \hline
    \end{tabular}
    \caption{Altitudes of detectors used, number of tracks and fat tracks observed, and expected amount of fat tracks based on Figure \ref{fig:altitude}.}
    \label{tab:altitudes}
\end{table}

The observed amount of fat tracks at each altitude matches the amount estimated by Figure~\ref{fig:altitude}, though further study is necessary to obtain more precise predictions and observations. The predictions are only an estimate based on Refs.~\citenum{CR-fluxes,muon-fluxes}, and the observed amounts are limited by the manner in which tracks are selected, which is highly dependent on the angle of particles incident on the detector. Streaks originate from particles incident at grazing angles, while particles which strike the detector too obliquely appear as spots instead, and thus are excluded from the analysis by the linear correlation coefficient filter. This means that the orientation of the CCDs during data collection could also affect the results. Since vertical cosmic rays are more common, following a distribution $\propto \cos^{2}\theta$ for muons and approximately $\propto \exp{(-1/\cos{\theta})}$ for nucleons \cite{ParticleReview}, CCDs perpendicular to the ground should be struck by more grazing particles and have a higher occurrence of streaks over spots than CCDs parallel to the ground. While the ComCam and PFS CCDs were perpendicular to the ground during observations, the LSST Camera CCDs were parallel to the ground, meaning they likely had proportionally fewer total cosmic rays which were included in our analysis. This study would benefit from both more careful data collection and more detailed calculations of atmospheric cosmic ray fluxes.

The distinction between muon and proton tracks in CCDs could be confirmed with a measurement of the masses of observed cosmic ray tracks. In principle, an estimate of the mass of a cosmic ray could be found by measuring the change in its energy deposition rate. However, due to high amounts of variation in $dE/dx$ in single tracks, we were unable to produce reliable mass estimates of any of our tracks, either fat or non-fat. Despite this, we believe that the increasing abundance of fat tracks with altitude does suggest protons as a strong candidate for the origin of fat tracks. 

Muons and protons could additionally be distinguished with the use of simulation tools such as \textsc{Geant4} \cite{Geant4}, which allows customized modelling of the passage of particles through matter. Such a study could be useful in determining the expected distributions of dE/dx for muons and protons, which could be directly compared to the distribution observed in Figure \ref{fig:dedxs}.

\section{Discussion and conclusion}
\label{sec:conclusion}

An alternate proposal for the method by which fat tracks occur is $\delta$ ray electrons, which are electrons which have been given enough energy by the primary cosmic ray to leave their own ionization trails. A large number of $\delta$ rays produced by the primary cosmic ray could have the effect of carrying energy away from the center of the track, producing a broadened image.

Ref.~\citenum{elec-ranges} calculates the continuous slowing down approximation (CSDA) range of a 10 keV electron in silicon to be about 1 \um{}. The maximum transferable kinetic energy to an electron from a proton with a velocity $\beta = 0.5$ (the average velocity of our measured fat tracks) is about 500 keV \cite{ParticleReview}. Thus the number of $\delta$ rays we expect to travel a non-negligible distance (i.e., greater than 1 \um{}) are those with energies between 10 and 500 keV.

Ref.~\citenum{ParticleReview} gives the distribution of $\delta$ rays per energy and distance. After integrating with respect to kinetic energy, we have \[\frac{dN}{dx} \approx \frac{1}{2} 4\pi N_A r_e^2 m_e c^2 z^2 \rho \frac{Z}{A} \frac{1}{\beta^2} (\frac{1}{10\ \rm{keV}} - \frac{1}{500\ \rm{keV}}) = 0.007\ \rm{\mu m}^{-1}\] for $\beta = 0.5$, or about one $\delta$ ray every 140 \um{} in silicon. As this is the rate at which all $\delta$ rays with ranges greater than 1 \um{} occur, we conclude that $\delta$ rays are incapable of producing the uniform broadening seen in fat tracks.

Although fairly uncommon, fat cosmic ray tracks have the potential to disrupt on-sky astronomical images, and we thus want to understand their causes in order to minimize their impact. We believe that fat tracks are most likely caused by proton cosmic rays, whose high energy deposition in CCDs results in significant electrostatic repulsion between electrons, broadening the resulting tracks in the image. Other mechanisms may contribute, but most of the measured widths of fat tracks can be explained by this phenomenon.

Because fat tracks come from grazing incident particles, we note that they are most likely to appear in detectors which are oriented perpendicular to the ground. Detectors at small angles relative to the ground should still see proton cosmic rays, but they should appear more like the round spots studied in the context of the plasma effect \cite{plasma-effect1,plasma-effect2,plasma-effect3}. While cosmic ray removal algorithms based on difference images are likely minimally impacted by fat tracks, those which rely on morphological identification of cosmic rays may need to consider how to remove tracks such as these, which do not match the typical narrow muon cosmic ray signature.

\section*{Code, Data, and Materials Availability}

The data utilized in this study were obtained from Vera C. Rubin Observatory and the Subaru Telescope. Data are available from the authors upon reasonable request, and with permission from their respective sources. The code used to generate results and figures is available at https://doi.org/10.5281/zenodo.8015806.

\section*{Acknowledgments}

This paper has undergone internal review in the LSST Dark Energy Science Collaboration. We thank the reviewers Pierre Astier, Zhiyuan Guo, and Craig Lage for their helpful comments. Additional thanks go to Merlin Fisher-Levine and Robert Lupton for their discussions.

This project was supported in part by the U.S. Department of Energy, Office of Science, Office of Workforce Development for Teachers and Scientists (WDTS) under the Science Undergraduate Laboratory Internships Program (SULI), and by the Brookhaven National Laboratory (BNL) Physics Department under the BNL Supplemental Undergraduate Research Program (SURP).

The DESC acknowledges ongoing support from the Institut National de 
Physique Nucl\'eaire et de Physique des Particules in France; the 
Science \& Technology Facilities Council in the United Kingdom; and the
Department of Energy, the National Science Foundation, and the LSST 
Corporation in the United States.  DESC uses resources of the IN2P3 
Computing Center (CC-IN2P3--Lyon/Villeurbanne - France) funded by the 
Centre National de la Recherche Scientifique; the National Energy 
Research Scientific Computing Center, a DOE Office of Science User 
Facility supported by the Office of Science of the U.S.\ Department of
Energy under Contract No.\ DE-AC02-05CH11231; STFC DiRAC HPC Facilities, 
funded by UK BEIS National E-infrastructure capital grants; and the UK 
particle physics grid, supported by the GridPP Collaboration.  This 
work was performed in part under DOE Contract DE-AC02-76SF00515.

Author contributions: TAG wrote and carried out the analysis code and wrote the paper. AN conceived the analysis idea, and contributed to the analysis and paper.

\bibliography{main}
\bibliographystyle{spiejour}

\textbf{Theodore Grosson} is currently pursuing their MSc degree in astronomy at the University of Victoria in British Columbia, Canada. Their research interest is in instrumentation for astronomy, particularly optical and infrared detectors, and they are currently developing new techniques for infrared spectrographs for their degree.

\textbf{Andrei Nomerotski} mostly worked on advanced detectors for High Energy Physics and Cosmology --- he is a strong believer that it is the new instrumentation and not new theories that lead to breakthroughs in science. Over the last two decades he went through different roles in several high profile institutions. He is currently working at Brookhaven National Laboratory (BNL) in the Uniteed States on production of the 3 Gigapixel digital camera for the next generation telescope, LSST. He is also leading another project at BNL (and previously at Oxford) to develop ultra-fast cameras for imaging 

\end{document}